\documentclass[english,aps,pra,twocolumn,amsmath,amssymb,superscriptaddress]{
revtex4-1}
\usepackage[latin9]{inputenc}
\usepackage{amstext}
\usepackage{amssymb}
\usepackage{graphicx}
\usepackage{color}
\usepackage{babel}



\begin{document}


\title{Sharp crossover from composite fermionization to phase separation in
mesoscopic mixtures of ultracold bosons}

\author{M. A. Garcia-March}
\email{magarciamarch@ecm.ub.edu}
\affiliation{Dept. d'Estructura i Constituents de la Mat\`eria, Univ. de
Barcelona, 08028 Barcelona, Spain}
  \author{B. Julia-Diaz}
\affiliation{Dept. d'Estructura i Constituents de la Mat\`eria, Univ. de
Barcelona, 08028 Barcelona, Spain}
\author{G.~E. Astrakharchik}
\affiliation{Dept. de F\'\i sica i Enginyeria Nuclear, Univ.  Polit\`ecnica de Catalunya, Barcelona, Spain}
\author{Th. Busch}
\affiliation{Quantum Systems Unit, Okinawa Institute of Science and
  Technology Graduate University, Okinawa, Japan}
\author{J. Boronat}
\affiliation{Dept. de F\'\i sica i Enginyeria Nuclear, Univ.  Polit\`ecnica de Catalunya, Barcelona, Spain}
  \author{A. Polls}
\affiliation{Dept. d'Estructura i Constituents de la Mat\`eria, Univ. de
Barcelona, 08028 Barcelona, Spain}

\begin{abstract}
We show that a two-component mixture of a few repulsively interacting ultracold atoms in a one-dimensional trap  possesses very different quantum regimes and that the crossover between them can be induced by tuning the interactions in one of the species. 
In the composite fermionization regime,  where the interactions between both components
are large, 
none of the species show large occupation of any natural orbital. 
Our results show that by increasing the interaction in one of the species, one can reach
the phase-separated regime.  In this regime, the weakly interacting component stays at the center
of the trap and becomes almost fully phase coherent, while  the strongly interacting component is displaced to the edges of the trap. 
The crossover is sharp, as observed in the in the energy and the in the largest occupation of a natural orbital of the weakly interacting species. 
Such a transition is a purely mesoscopic effect which disappears for large atom numbers. 
\end{abstract}

\maketitle


Ensembles of a few interacting ultracold trapped atoms constitute  unique quantum systems. They can be exceptionally well isolated from the
environment, minimizing the role of decoherence and are perfect candidates to implement states with strong quantum correlations. Moreover, they are extremely
versatile,   as precise control over both the shape of the trapping potential and the 
atom-atom interactions is routinely realized in currents experiments. Also, all their degrees of freedom other than the positions can be ignored, thus providing {\it simple} systems which  nevertheless show a
great diversity of phenomena~\cite{Blume2012,Bloch2008}. 
Experimentally, 
a system of small number of strongly repulsive bosons, that is the Tonks-Girardeau (TG) gas, has been realized
both in a single trap and in an optical lattice
(OL)~\cite{Kinoshita2004}. Furthermore, the loading of a small number of atoms  in a single
trap, both for fermionic and bosonic
species, has been achieved~\cite{He:10}, and a great experimental effort
in this direction is undertaken in laboratories  worldwide, a further step being the realization of  few-atom bosonic mixtures in a single trap.

\begin{figure}
\begin{tabular}{cc}
\includegraphics[scale=0.3]{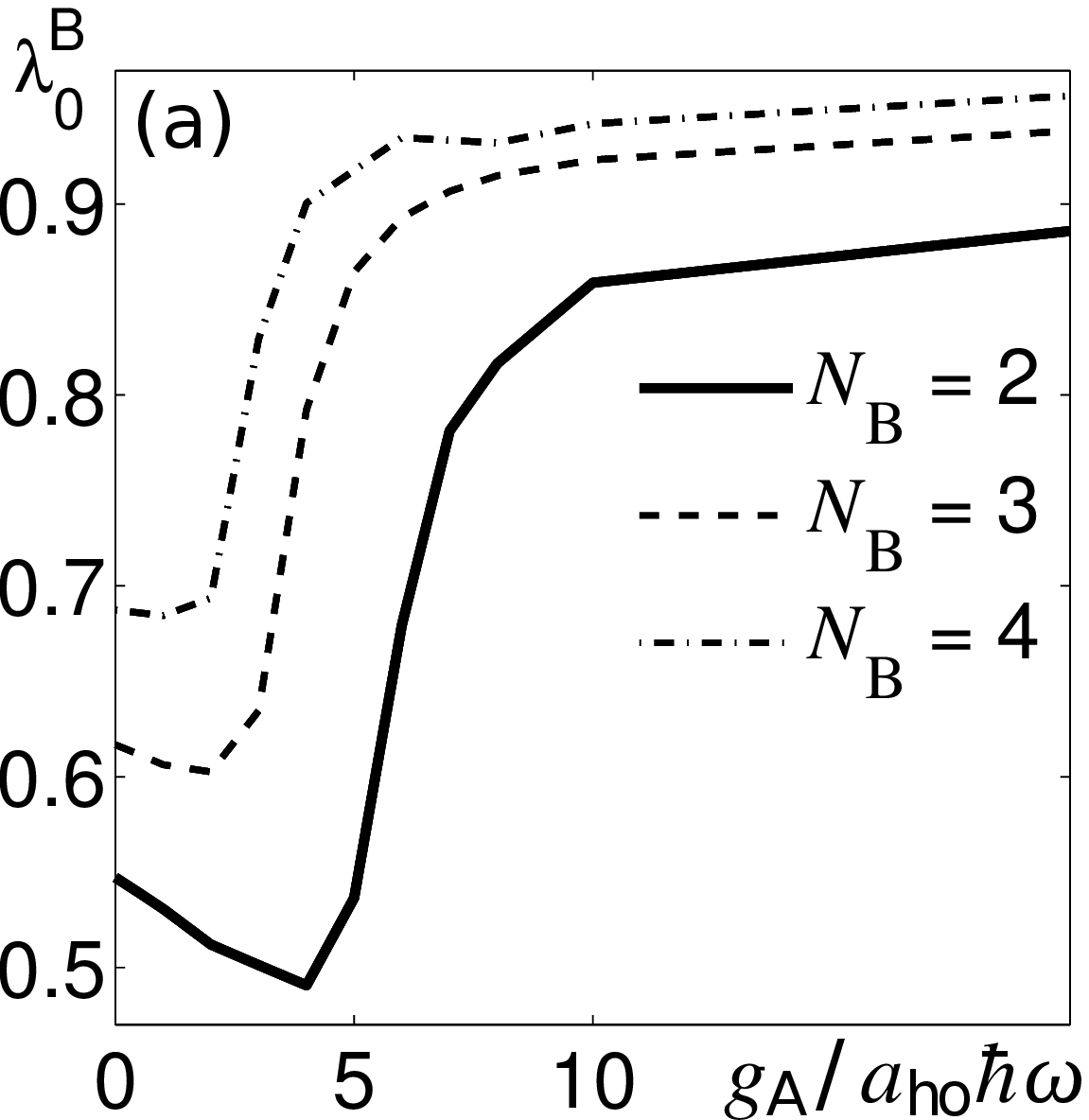}  &
\includegraphics[scale=0.29]{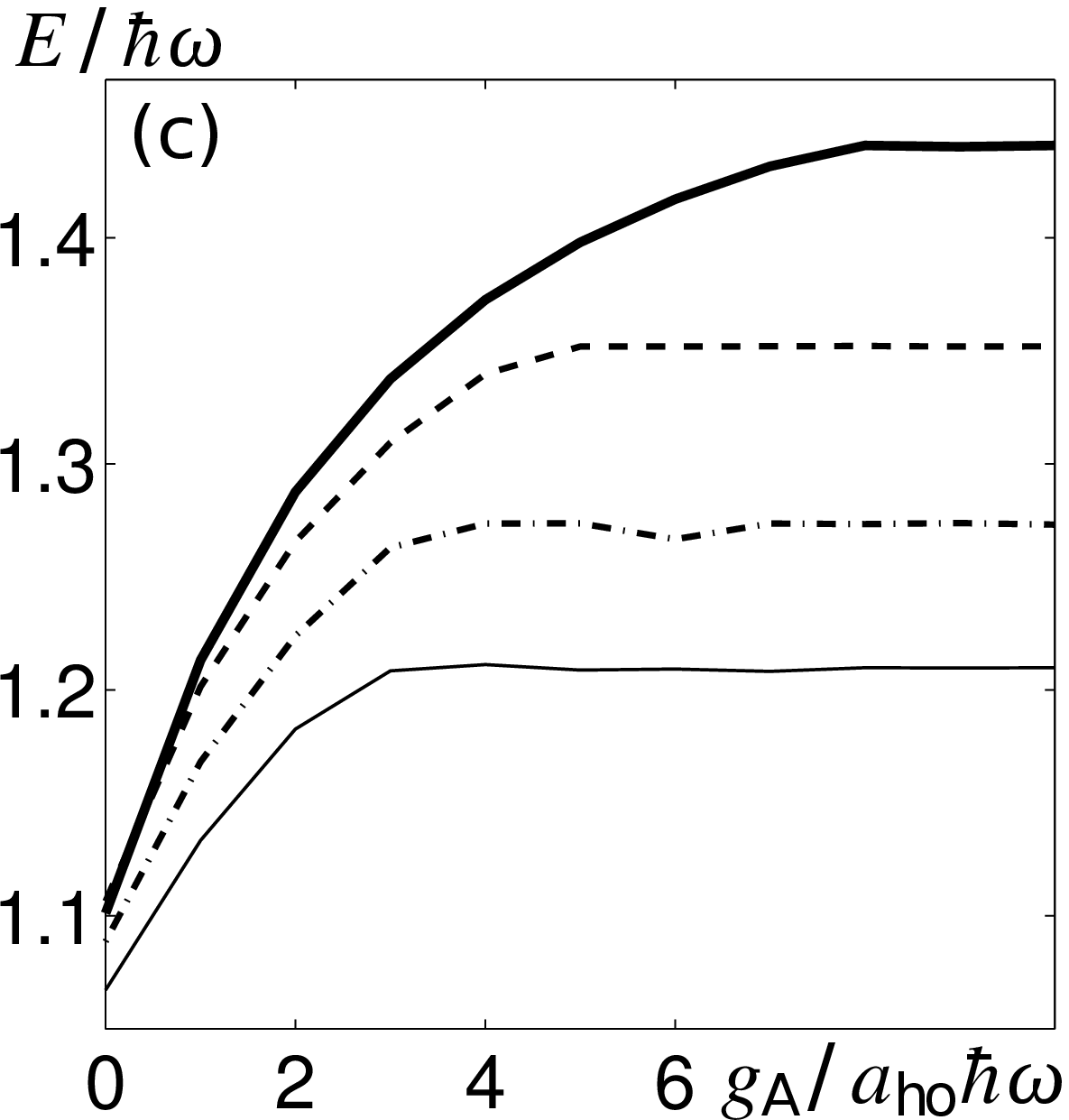} \\
\includegraphics[scale=0.3]{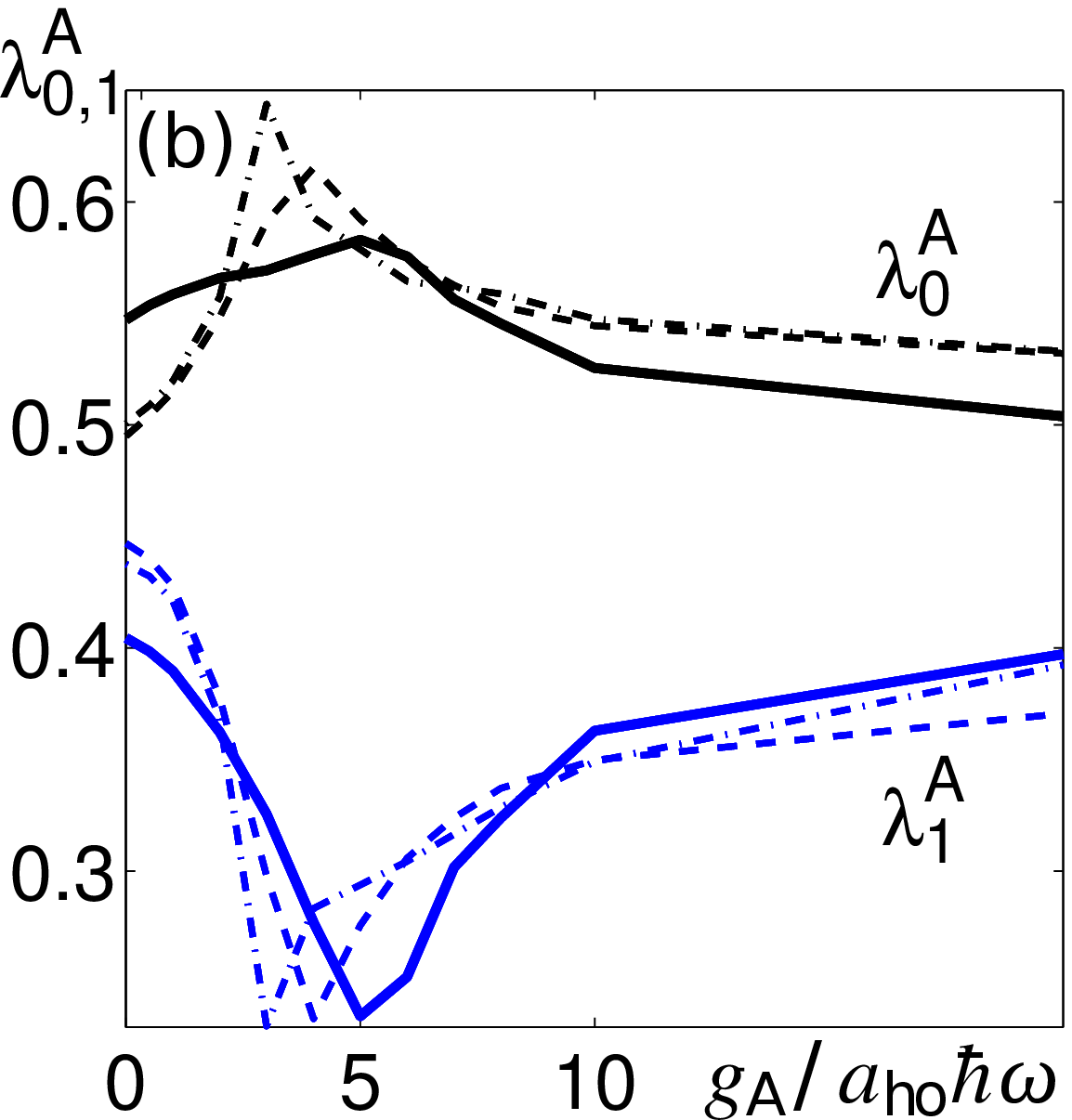}  &\hspace{0.05cm}
\includegraphics[scale=0.3]{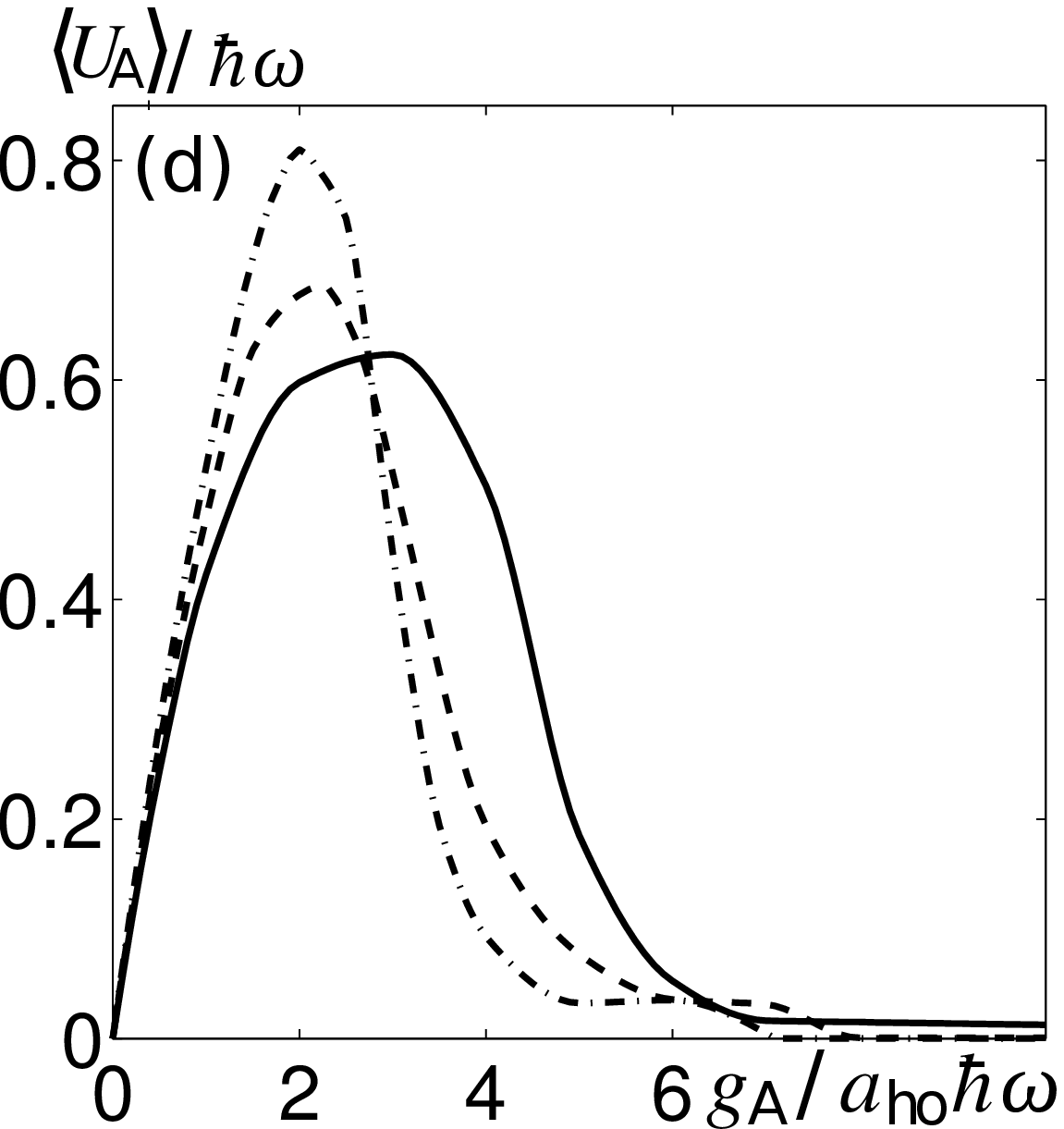} \vspace{-0.25cm}
\end{tabular}
\caption{(Color online) {\it Largest occupations $\lambda_0$ and $\lambda_1$ of the natural orbitals, energy per atom $E$, and interaction energy $\langle U_{\mathrm{A}}\rangle$. } (a) largest occupation number $\lambda_0$ for species B, for  $N_{\mathrm{A}}\!=\!2$, and $N_{\mathrm{B}}\!=\!2,3,4$   calculated with direct diagonalization. (b) largest occupation number of a natural orbital for species A (black upper lines) and occupation of the second highest occupied orbital (blue bottom lines). (c)  energy per atom for $N_{\mathrm{B}}\! =\! 2,3,4,5$   calculated using DMC (solid thin line for $N_{\mathrm{B}}\!=\!5$). (d)  average interaction energy of species A, $\langle U_{\mathrm{A}}\rangle\! =\!\langle\sum_{j<j'}^{N_{A}}\!v_{\mathrm{int}}^A(x_j,x_{j'}) \rangle$. In all cases, $g_{\mathrm{B}}\! =\! 0$ and $g_{\mathrm{AB}}$ is large. \vspace{-0.75cm}
} \label{fig1}
\end{figure}

The experimental loading of more than one boson in a small microtrap is hindered by
 the  collisional blockade mechanism~\cite{Schlosser2002}. For example, it is possible to load a single atom   
 in a dipole trap from  cold bosons trapped in a magneto optical
trap~\cite{Schlosser2001} or in an
OL~\cite{Orzel2001}. Double, or higher  occupancy can be realized in the Mott insulating phase in an OL combined with a parabolic trap~\cite{Greiner2002}. These systems can be used to implement schemes of quantum computation, if
combined with single atom
detection~\cite{Bloch2008,Nelson2007}, where
multiple occupation of single sites can also be 
resolved~\cite{Bakr2010}. Other examples of few ultracold atom experiments are the TG
gas~\cite{Kinoshita2004} and the recent realizations mentioned
above~\cite{He:10}. 
 These systems are a natural ground for studies in squeezing and entanglement
with applications, for example, in precision
measurements~\cite{Orzel2001,Esteve2008}, thus leading to the great interest in
their experimental realization. 


 In this Letter, we predict a sharp crossover between two very different
regimes in mesoscopic mixtures of repulsively interacting bosons (species A and B), corresponding to
different interaction strengths  between atoms of same and different species.
In one of the limiting regimes, the interactions between the atoms of
{\it different} species are strong, while the interactions between the atoms of
the {\it same} species are very weak. This was termed as the {\it composite
fermionization} limit~\cite{Zollner2008a,Alon:2007}. It has some common features with a
TG  gas~\cite{Girardeau2001}, particularly the presence of zeros in the
wavefunction whenever two atoms of different species meet. The largest occupation of a natural orbital scales with number of particles $N$ faster than $\sqrt N$, but slower than $N$. Therefore, the system is not fully Bose-Einstein condensed. 
Another limit is obtained when the interactions A-A in  species A are as  large as the  interactions between  both species, A-B,  keeping the interactions B-B between
the atoms of species B small. In this second limit, phase separation
occurs and the occupation of a single natural orbital of species B tends to $N$. Therefore, species B condenses in the center of the trap. As we will show,
the transition between both limits is sharp, while both limits are clearly
distinguishable  in terms of both the one- and two-body correlations, and in the degree of
condensation of B. Along this process, a highly quantum correlated state is revealed, which is produced by increasing the A-A interactions.

We consider a small number of atoms  of two bosonic components 
trapped in a one-dimensional harmonic potential
$V(x)=\frac{1}{2}m\omega^{2}x^{2}$. We suppose that each component is a different hyperfine state of the same atomic species, and therefore the atoms of  both components have the same mass $m=m_{\mathrm{A}}=m_{\mathrm{B}}$.   
The short-range interactions in each species is described by a
contact $\delta$-potential $v_{\mathrm{int}}^{\mathrm{A}}=g_{\mathrm{A}}\delta(x_j-x_{j'}) $ and
$v_{\mathrm{int}}^{\mathrm{B}}=g_{\mathrm{B}}\delta(y_j-y_{j'}) $, with $x_j$ ($y_j$) standing for the
position of atom $j$ in species A (B).  
Atoms of different species interact also by a $\delta$-potential
$v_{\mathrm{int}}^{\mathrm{AB}}=g_{\mathrm{AB}}\delta(x_i-y_{j}) $. Here $g_{\mathrm{A (B)}}$ and $g_{\mathrm{AB}}$
are the one-dimensional intra- and interspecies coupling constants, respectively. We assume that all constants are positive corresponding to repulsive interactions. These coupling constants can be finely tuned by means of Feshbach resonances and confinement induced resonances~\cite{Olshanii1998}.  
The  Hamiltonian for  fixed number of atoms $N_{\mathrm{A,B}}$ in each species
can then be written as $\hat H= \hat H_{\mathrm{A}}+\hat H_{\mathrm{B}} +\hat H_{\mathrm{AB}} $, where
\begin{align}
 &\hat H_{\mathrm{A}}=\sum_{j=1}^{N_{\mathrm{A}}}\left[\frac{-\hbar^2}{2m}\frac{\partial^2}{\partial
x_j^2}+V(x_j)
\right]+\sum_{j<j'}^{N_{\mathrm{A}}}v_{\mathrm{int}}^{\mathrm{A}}(x_j,x_{j'}),\nonumber\\
 &\hat H_{\mathrm{B}}=\sum_{j=1}^{N_{\mathrm{B}}}\left[\frac{-\hbar^2}{2m}\frac{\partial^2}{\partial
y_j^2}+V(y_j)
\right]+\sum_{j<j'}^{N_{\mathrm{B}}}v_{\mathrm{int}}^{\mathrm{B}}(y_j,y_{j'}),\nonumber\\
 &\hat H_{\mathrm{AB}} =\sum_{j=1}^{N_{\mathrm{A}}} \sum_{j'=1}^{N_{\mathrm{B}}}
v_{\mathrm{int}}^{\mathrm{AB}}(x_j,y_{j'}).  
\end{align}
It is natural to use harmonic oscillator units, that is to scale all lengths in terms of the oscillator length $a_{\mathrm{ho}}=\sqrt{\hbar/(m\omega)}$ and all energies in units of $\hbar\omega$. 
If one of the coupling constants becomes large, the wave
function vanishes when  two of the interacting atoms meet.

We propose the following {\it ansatz} to describe the physics of the mixture
\begin{align}
\label{wavefunc1}
 &\Psi(X,Y)=\Phi(X) \, \Phi(Y)
 \prod_{j<k}^{N_{\mathrm{A}}}|x_k-x_j-a_{\mathrm{A}}|\nonumber\\
& \prod_{j<k}^{N_{\mathrm{B}}}|y_k-y_j-a_{\mathrm{B}}|\prod_{j,k}^{N_{\mathrm{A}},N_{\mathrm{B}}}|x_k-y_j-a_{\mathrm{AB}}|,
\end{align} 
 where the Gaussian function $\Phi$ is  the exact solution for non-interacting atoms in the  harmonic trap, $\Phi(X)= \exp [-\sum x_i^2 / (2 a_{\mathrm{ho}}^2) ]$ and $X=\{x_i\}$ and $Y=\{y_i\}$.   
 The one-dimensional $s$-wave scattering length $a_{\sigma}$,  with $\sigma=\mathrm{A,B, AB}$ for the interactions between A-A, B-B, and A-B atoms, is related to the corresponding coupling constant as $g_{\sigma}=-2\hbar^2/(m a_{\sigma})$. 
The crossover discussed in this Letter occurs for ideal Bose-gas interactions in species B, $g_{\mathrm{B}}=0$, and hard-core TG A-B interactions, $g_{\mathrm{AB}}\rightarrow\infty$, when the  interactions A-A are tuned from ideal to TG gas. In this situation the wavefunction can be simplified to 
\begin{align}
\label{wavefunc2}
 \Psi=\Phi(X) \, \Phi(Y)
 \prod_{j<k}^{N_{\mathrm{A}}}\!|x_k\!-\!x_j\!-\!a_{\mathrm{A}}|
\!\!\prod_{j,k}^{N_{\mathrm{A}},N_{\mathrm{B}}}\!\!|x_k\!-\!y_j|. 
\end{align}

In the limit of vanishing $g_{\mathrm{A}}$,  the terms containing $a_{\mathrm{A}}\rightarrow\infty$ drop out from Eq.~(\ref{wavefunc2}) and the wavefunction only has nodes whenever two atoms of different species meet, which corresponds to the  composite fermionization limit~\cite{Zollner2008a}. 
%
On the other hand, the wavefunction~(\ref{wavefunc2}) for hard-core A-A interactions ($g_{\mathrm{A}}\rightarrow\infty$ and $a_{\mathrm{A}}=0$),  gains additional nodes and vanishes whenever two atoms of species A-A or A-B meet. 
The system is a mixture of a TG gas and an ideal gas, which we coin as TG-BEC gas. In these limits the wavefunction~(\ref{wavefunc2}) is not exact, but   it  correctly describes the physical properties, as we will discuss below. Finally, if all coupling constants are large the system falls within a family of soluble models  discussed in Ref.~\cite{Girardeau2007}.

Here we are mainly interested in the transition from
composite fermionization to a  TG-BEC gas as   $g_{\mathrm{A}}$ is tuned from zero to large
values. The TG-BEC gas for $N_{\mathrm{B}}\ge N_{\mathrm{A}}$ is
spatially separated~\cite{Garcia-march2012}. Then, the occupation of a single natural orbital for species B is of the order of $N_{\mathrm{B}}$, and it is Bose-Einstein condensed in the center of
the trap while A occupies the outer region of the trap. Conversely,
in the composite fermionization limit the occupation of a natural orbital for both species is smaller than $N_{A,B}$, and therefore none of the species is fully Bose-Einstein condensed. Our main goal is to characterize the transition between these two very different limits as we increase $g_{\mathrm{A}}$. 

We calculate the ground state  by direct diagonalization of the Hamiltonian after expressing the field operators $\hat{\Psi}(x)$ in terms of the single particle eigenfunctions \cite{Garcia-march2012}.  From this we obtain the one-body density matrix (OBDM), $\rho_1=\langle\hat{\Psi}^\dagger(x)\hat{\Psi}(x')\rangle$, and its diagonalization gives  the natural
orbitals and their occupation numbers $\lambda_i$ (with $\lambda_0$ being the largest of them).  
In Fig.~\ref{fig1} (a) and (b), we show the largest occupation numbers of the
natural orbitals of the ground state as we increase $g_{\mathrm{A}}$, with  $g_{\mathrm{B}}=0$, and $g_{\mathrm{AB}}=500\hbar\omega a_{\mathrm{ho}}$, when $N_{\mathrm{A}}=2$ and
$N_{\mathrm{B}}=2,3,4$. 
The largest
occupation for the ideal Bose gas B,  $\lambda_0^{\mathrm{B}}$, shows a sharp crossover between the composite
fermionization and the phase-separated TG-BEC limits, while $\lambda_0^A$ is reduced to $\sim 0.5$. 
This crossover also corresponds to an increase in energy until a plateau is reached see Fig.~\ref{fig1} (c), where the energy per particle for all cases in (a) is shown obtained from direct diagonalization and  from a Diffusion Quantum Monte Carlo
(DMC) method using the wavefunction~(\ref{wavefunc2}) as a guiding function. The energies calculated with direct diagonalization are compatible with the  ones calculated with DMC~\cite{Boronat1994}. We observe that in all cases a plateau is reached for a characteristic $g_{\mathrm{A}}$, whose value becomes smaller as  $N_{\mathrm{B}}$  is increased. 

\begin{figure}
\begin{tabular}{ccc}
 \hspace{-0.15cm}\includegraphics[scale=0.6]{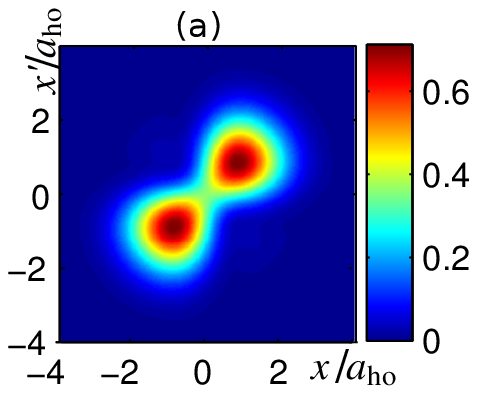} &\hspace{-0.1cm}
 \includegraphics[scale=0.6]{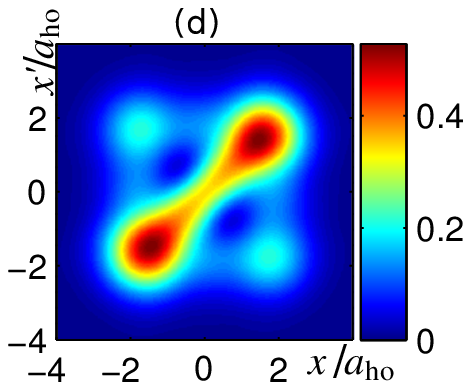}  &\hspace{-0.1cm}
 \includegraphics[scale=0.6]{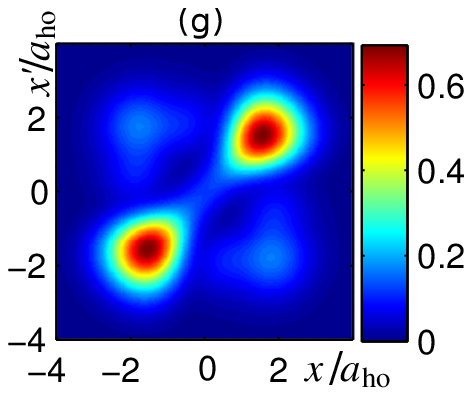} \\
 \hspace{-0.15cm}\includegraphics[scale=0.6]{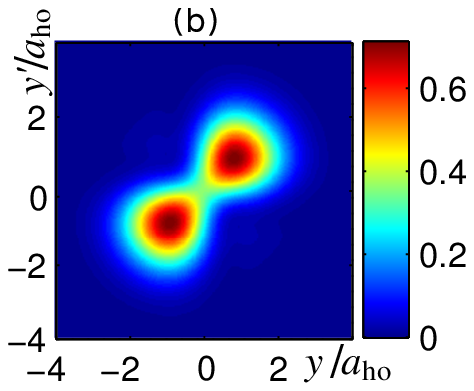} &\hspace{-0.1cm}
 \includegraphics[scale=0.6]{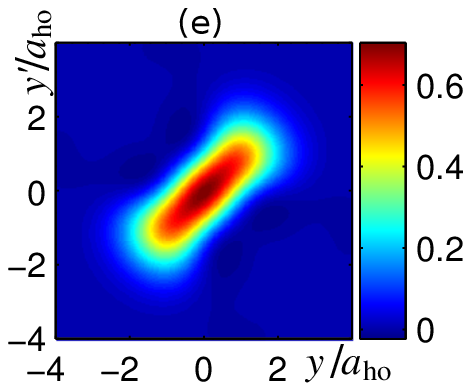}  &\hspace{-0.1cm}
 \includegraphics[scale=0.6]{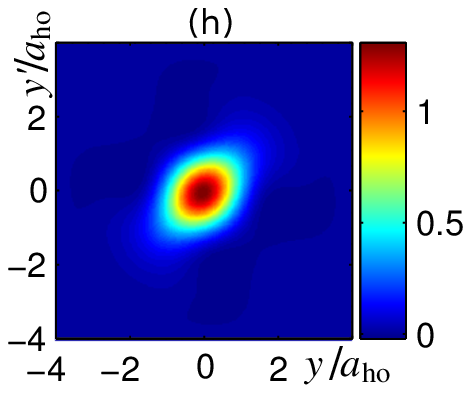} \\
 \hspace{-0.55cm}\includegraphics[scale=0.7]{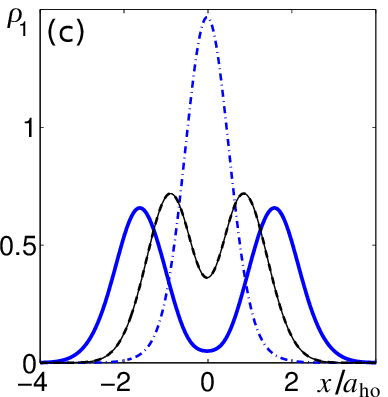} &\hspace{-0.45cm}
 \includegraphics[scale=0.7]{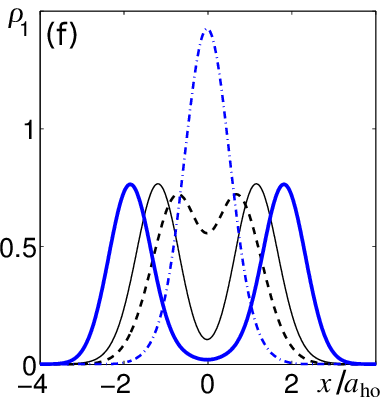}  &\hspace{-0.45cm}
 \includegraphics[scale=0.7]{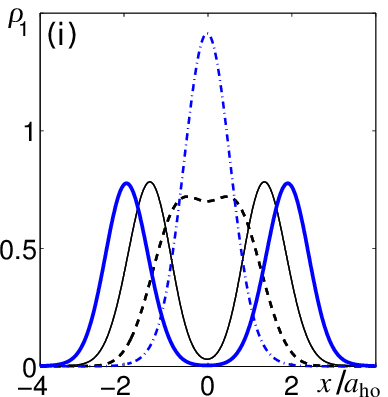} 
\end{tabular}
\caption{(Color online) {\it Contour plot of the one-body density matrix $\rho_1(x,x') $  in the composite fermionization and the  phase-separated TG-BEC   limits.  } Here $N_{\mathrm{A}}=N_{\mathrm{B}}=2$,  $g_{\mathrm{B}}=0$, and $g_{\mathrm{AB}}$ is large.
Panels (a) and (b) show  the OBDM for species A and B, respectively, in the composite fermionization limit
($g_{\mathrm{A}}=0$); Panels (d) and (e) depict the intermediate case, $g_{\mathrm{A}}=5\hbar\omega a_{\mathrm{ho}}$ and panels (g) and (h) the phase-separated TG-BEC limit, $g_{\mathrm{A}}=7\hbar\omega a_{\mathrm{ho}}$.    Panels  (c), (f), and (i) show the density profile in the composite fermionization limit [$g_{\mathrm{ho}}=0$, black solid thin line for A, dashed line for B] and in the phase-separated TG-BEC limit  [$g_{\mathrm{ho}}=10\hbar\omega a_{\mathrm{ho}}$, blue  solid thick line for A, dash-dotted line for B], for $N_B=2,3,4$ respectively.    \label{fig2}}
\end{figure}

To understand the crossover behavior observed in Fig.~\ref{fig1}, let us first discuss the behavior of the correlations present in $\rho_1$ and $\rho_2$. The two-body distribution
function (TBDF),   $\rho_2$, for two atoms of  the same species is defined as
\begin{align*}
&\rho_2^{\mathrm{A}}(x_1,x_2)\!=\!N_{\mathrm{A}}(N_{\mathrm{A}}\!-\!1)\!\int\! dx_3\cdots dx_{N_{\mathrm{A}}} dy_1\cdots dy_{N_{\mathrm{B}}}\,|\Psi|^2, 
\end{align*}
and  the  cross two-body distribution function (CTBDF)  for two atoms of  different
species is
\begin{align*}
&\rho_2^{\mathrm{AB}}(x_1,y_1)\!=\!N_{\mathrm{A}}\,N_{\mathrm{B}}\!\int\! dx_2\cdots dx_{N_{\mathrm{A}}}
dy_2\cdots dy_{N_{\mathrm{B}}}\,|\Psi|^2.
\end{align*}
The TBDF for A (B) represents the probability of finding an atom of A (B)
at $x_2\, (y_2)$  when one atom of the same species has been found at $x_1\,
(y_1)$. The CTBDF is the
probability of finding an atom of B  at $y_1$ when one atom of A   has been
found at $x_1$. 

\begin{figure}
\begin{tabular}{ccc}
 \hspace{-0.15cm}\includegraphics[scale=0.6]{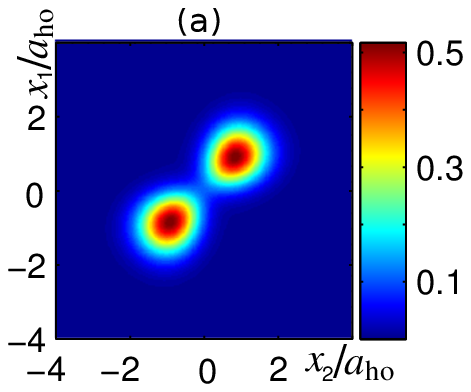} &\hspace{-0.1cm}
 \includegraphics[scale=0.6]{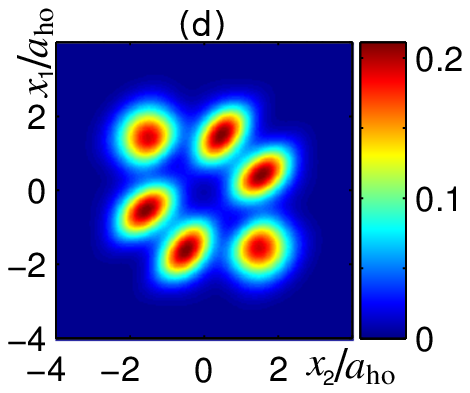}  &\hspace{-0.1cm}
 \includegraphics[scale=0.6]{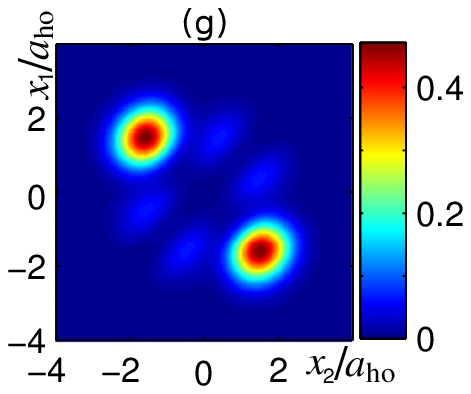} \\
 \hspace{-0.15cm}\includegraphics[scale=0.6]{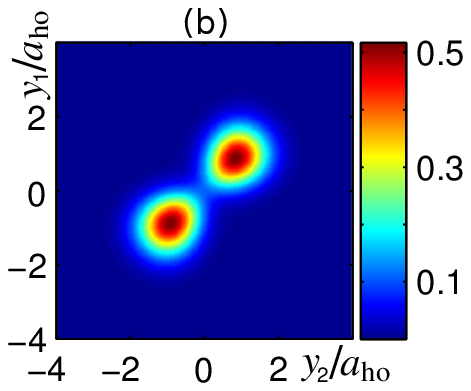} &\hspace{-0.1cm}
 \includegraphics[scale=0.6]{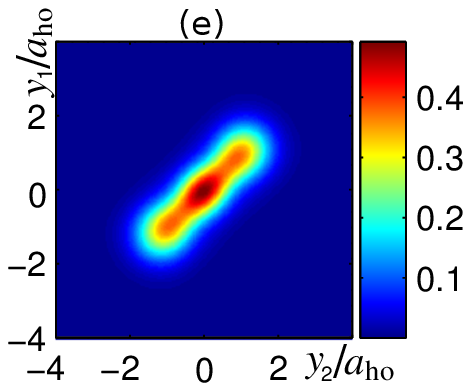}  &\hspace{-0.1cm}
 \includegraphics[scale=0.6]{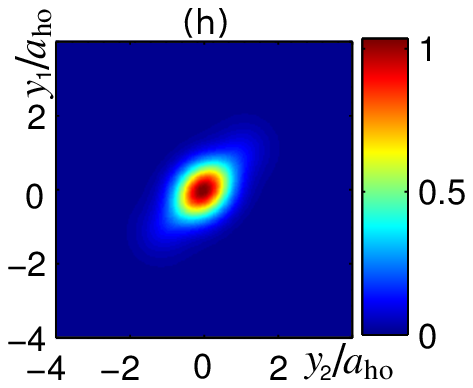} \\
 \hspace{-0.15cm}\includegraphics[scale=0.6]{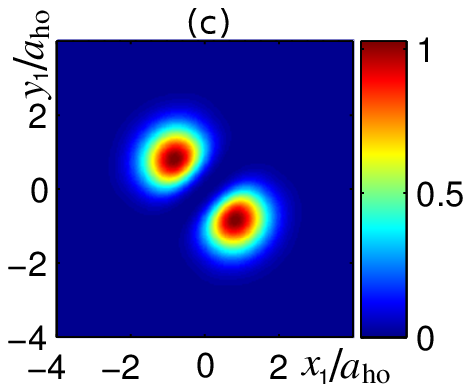} &\hspace{-0.1cm}
 \includegraphics[scale=0.6]{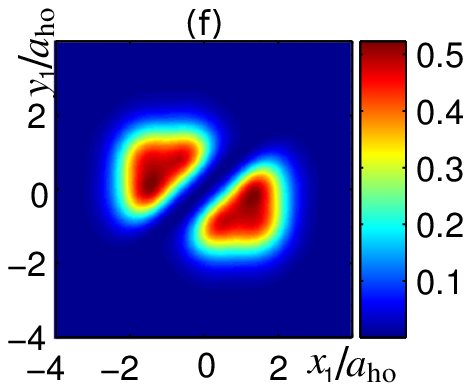}  &\hspace{-0.1cm}
 \includegraphics[scale=0.6]{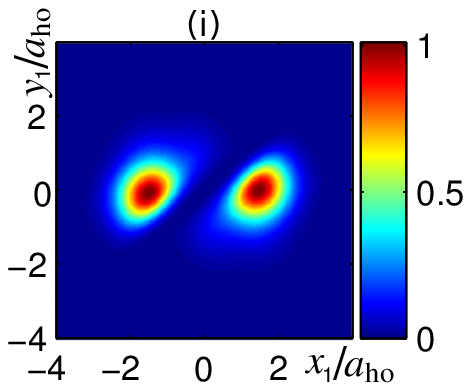} 
\end{tabular}
\caption{(Color online) {\it Contour plot of the two-body distribution functions in the composite fermionization and the phase-separated TG-BEC limits  for the same parameters as in Fig.~\ref{fig2}. 
}   (a)  and (b) show the TBDF for species A and B, and (c) the CTBDF in the
composite fermionization limit ($g_{\mathrm{A}}=0$); (d)-(f) show the same for the intermediate limit,  $g_{\mathrm{A}}=5\hbar\omega a_{\mathrm{ho}}$ and  (g)-(i) for the  phase-separated TG-BEC limit, $g_{\mathrm{A}}=7\hbar\omega a_{\mathrm{ho}}$.    \label{fig3}}
\end{figure}

In the composite fermionization limit (e.g. for $g_{\mathrm{A}}=g_{\mathrm{B}}=0$ and $g_{\mathrm{AB}}=500\hbar\omega a_{\mathrm{ho}}$), the OBDMs for both species are identical when $N_{\mathrm{A}}=N_{\mathrm{B}}$, as illustrated in Figs.~\ref{fig2} (a) and (b). The diagonal elements  $x=x' (y=y')$ of the OBDM correspond to the density profile  [see Fig.~\ref{fig2} (c)].  The
largest occupation of a natural orbital is  significantly depleted,  $\lambda_0^{A,B}\simeq0.55$ [see
Fig.~\ref{fig1} (a) and (b)]. 
The corresponding TBDFs for A and B species
are represented in Figs.~\ref{fig3} (a) and (b) respectively. As one can see,  two atoms of the same species can occupy  the same position. Conversely,  two atoms of different species avoid each other, as manifested in the  CTBDF depicted in Fig.~\ref{fig3} (c), which  vanishes along the  $x_1=y_1$ diagonal. Then, if an atom of species A is found in one
of the regions where the density has a peak, all atoms of species B will be found  in 
the region
corresponding to the other peak.    
We notice that the wavefunction~(\ref{wavefunc2}) is  a good approximation, as it correctly reproduces the density and TBDFs (not shown). In general, the same behavior of the density and TBDFs is expected whenever $N_{\mathrm{A}}=N_{\mathrm{B}}>2$, if the interspecies interactions are large enough to produce the zeros present in wavefunction~(\ref{wavefunc2}).   
In an unbalanced mesoscopic system,  $N_{\mathrm{B}}>N_{\mathrm{A}}$, the OBDM  is
not equal for both species [see Figs.~\ref{fig2} (f) and (i)], as species B has a greater tendency to occupy the
center of the trap, but it is still not fully condensed, as can be seen from
Fig.~\ref{fig1}(a). In this case the system remains in the composite fermionization limit.

In the TG-BEC limit, phase separation occurs. The spatial overlap of both species is significantly reduced, as can be seen in the density profiles of both species, i.e. the diagonal elements of the OBDMs (Fig.~\ref{fig2} (g) and (h) for  $g_{\mathrm{A}}=7 \hbar\omega a_{\mathrm{ho}}$). The occupation of the lowest natural
orbital of species B is high, $\lambda_0^{\mathrm{B}} \simeq0.9$, indicating that this
species is almost fully condensed. On the contrary,  the largest occupation of a natural orbital for A drops down,  $\lambda_0^{\mathrm{A}} \simeq0.5$, as this
species is fragmented into two  parts, each located at either side of B (see Fig.~\ref{fig1}(b)). The
TBDF for B corresponds to a condensed cloud in the center of the trap (Fig.~\ref{fig3}(h)), while
the one of A corresponds to a fragmented gas of two single atoms at each side
of B (Fig.~\ref{fig3}(g)).   The CTBDF (see Fig.~\ref{fig3}(i)) shows
that the atoms of species B will be found at the center of the trap with the
largest probability, while the atoms of species A stay at the edges of the trap. 

For  $g_{\mathrm{A}}$ very close to the crossover, the density profiles of both components still overlap (see, for example, Figs.~\ref{fig2} (d) and (e), where $g_{\mathrm{A}}=5\hbar\omega a_{\mathrm{ho}}$). In this intermediate regime, the occupation of a single natural orbital of species B gets depleted with increasing A-A interaction strength until a certain minimum is reached. Then, it rapidly grows until the phase-separated  limit is reached, in which the condensation of this species is complete (see Fig.~\ref{fig1}(a)). 
The quantum correlations are enhanced as the crossover is approached. As the interaction strength in A is increased, the two atoms in A tend to avoid each other. This induces a minimum along the diagonal $x_1=x_2$ in the TBDF for species A (see Fig.~\ref{fig3}(d)). On top of this, phase separation is favored, and therefore the probability of finding an atom of species A at each side of the trap grows (see peaks along diagonal $x_1=-x_2$ in Fig.~\ref{fig3}(d)). The TBDF of species B  (see Fig.~\ref{fig3} (e)) shows a maximum in the center of the trap, where the atoms in this species tend to locate. The interactions between both species are strong, and therefore the CTBDF (Fig.~\ref{fig3} (f)) keeps the zero along the diagonal $x_1=y_1$. This CTBDF also has a non-zero value close to $y_1=0$ because the atoms of B tend to locate at the center of the trap. 
Therefore, the TBDF for A indicates that two atoms of A cannot be found
at the same position of space, corresponding to a TG gas, but they can be found
either to the right or to the left of the atoms of B, which are located in the center
of the trap.  This corresponds to a highly correlated mesoscopic superposition
state where the two atoms of A are on the left and on the right of the atoms of B, which is induced in the system by increasing the interaction strength in A. This can be a new route to the experimental creation of superposition states, which have potential applications in quantum information protocols.
These one- and two-body correlation functions are well reproduced by using wavefunction Eq.~(\ref{wavefunc2}). 

When the population in species B is increased, the crossover occurs at smaller
values of $g_{\mathrm{A}}$. To show
this, we plot in Fig.~\ref{fig1} (c) the cases  $N_{\mathrm{B}}=2,3,4,5$ .  For  $g_{\mathrm{A}}$ small, the growth of the energy is similar for all cases.  
Indeed, for  $ g_{\mathrm{A}}=0$,  the condensation of species B becomes complete for large $N_{\mathrm{B}}$ (see Fig.~\ref{fig1} (a)), indicating that this crossover only appears for mesoscopic unbalances in the population of both components.

Finally, the position of the sharp crossover can also be clearly observed in the interaction energy $\langle U_{\mathrm{A}} \rangle$. It vanishes both in the composite fermionization limit, because $g_{\mathrm{A}}=0$ and in the TG-BEC limit, because $ \rho_2^{\mathrm{A}}(x_1,x_1)=0$ (see Fig.~\ref{fig1}(d)). In-between both limits, the interaction energy for A shows a maximum which approximately coincides with the position of the crossover.  In order to find the order of the transition, at zero temperature one has to study the continuity of  $dE/d g_{\mathrm{A}}$. 
By means  of the Hellmann-Feynman theorem one finds that $g_{\mathrm{A}} dE/d g_{\mathrm{A}}=\langle\psi_{\mathrm{gs}}|U_{\mathrm{A}}|\psi_{\mathrm{gs}}\rangle$. Since  $\langle U_{\mathrm{A}} \rangle$ is continuous, the transition is a crossover. 

In conclusion, by using a diagonalization method and a DMC approach, we have studied in an exact way the ground state properties of a small, trapped one-dimensional two-component Bose gas. We have identified the regimes of composite fermionization and phase separation, and we find a sharp crossover between them. In the phase separation regime, one of the components gets fragmented into separated pieces, each of them preserving coherence.  
We believe that this new crossover is of key interest for understanding upcoming experiments with small mixtures of interacting ultracold
bosons, and the macroscopic superposition state shown to exist between both regimes has potential outreach to applications on quantum information. We leave  for future
research those cases in which the number of atoms in the species with tunable
interactions is higher than the number of atoms in the weakly interacting one, as in
this case the latter is not necessarily the one that occupies the center of the trap.   

\begin{acknowledgments}
  This project was support by Science Foundation Ireland under Project
  No.~10/IN.1/I2979. We acknowledge also partial financial support from the DGI (Spain) Grant No.
FIS2011-25275, FIS2008-00784 (TOQATA), and  the Generalitat de Catalunya Grant No. 2009SGR-1003. GEA and BJD acknowledge fellowship by MEC (Spain) through the Ramon y Cajal program.  
\end{acknowledgments}

\end{document}